\documentclass{emulateapj}

\usepackage{natbib}
\usepackage{amsmath}
\usepackage{lscape}
\usepackage{longtable}

\newcommand{\about}{$\sim\!\!$~}

\newcommand{\be}{\begin{displaymath}}
\newcommand{\ee}{\end{displaymath}}

\def\lsim{\hbox{\rlap{\raise 0.425ex\hbox{$<$}}\lower 0.65ex\hbox{$\sim$}}}
\def\gsim{\hbox{\rlap{\raise 0.425ex\hbox{$>$}}\lower 0.65ex\hbox{$\sim$}}}

\shorttitle{UV Spectra of SNe~Ia}
\shortauthors{Foley, Filippenko, \& Jha}

\begin{document}

\title{Luminosity Indicators in the Ultraviolet Spectra of Type Ia Supernovae}

\author{
{Ryan~J.~Foley,}
{Alexei~V.~Filippenko}
}

\affiliation{Department of Astronomy, University of California,
Berkeley, CA 94720-3411 (rfoley@astro.berkeley.edu,
alex@astro.berkeley.edu)}
\and

\author{{Saurabh~W.~Jha}}

\affiliation{Dept. of Physics and Astronomy, Rutgers, the State
University of New Jersey, 136 Frelinghuysen Road, Piscataway, NJ 08854
(saurabh@physics.rutgers.edu)}

\begin{abstract}
We present a complete sample of {\it International Ultraviolet
Explorer} and {\it Hubble Space Telescope} ultraviolet (UV) spectra of
Type Ia supernovae (SNe~Ia) through 2004.  We measure the equivalent
width (EW) and blueshifted velocity of the minimum of the one strong
UV feature, \ion{Fe}{2} $\lambda 3250$.  We also quantify  the slope
of the near-UV spectra using a new parameter, the ``UV ratio.''  We
find that the velocity of the \ion{Fe}{2} line does not correlate with
light-curve shape, while the EW shows distinct behavior for the slow
and fast-declining objects.  Using precise Cepheid and surface
brightness fluctuation distance measurements of 6 objects with
UV spectra observed near maximum light (a total of 12 spectra), we
determine that the UV ratio at maximum light is highly correlated with
SN~Ia luminosity.  A larger sample of UV spectra is necessary to
determine the validity of these luminosity indicators and whether they can
be combined with light-curve shape to improve measured SN~Ia distances.

\end{abstract}


%%%%%%%%%%%%%%%%%%%%
%%  Introduction  %%
%%%%%%%%%%%%%%%%%%%%

\section{Introduction}\label{s:intro}

Ultraviolet (UV) spectra of Type Ia supernovae (SNe~Ia) have been
incredibly helpful in understanding the explosion physics of SNe~Ia
\citep[e.g.,][]{Jeffery92}; over the last 16 years, however, very few
such spectra have been obtained.  Compounding the problem, the demise
in 2004 of the Space Telescope Imaging Spectrograph (STIS) aboard the
{\it Hubble Space Telescope} ({\it HST}) has stalled progress in this
field.

In addition to providing a better understanding of SN~Ia explosions,
UV spectra are critical for properly measuring luminosity distances of
high-redshift SNe~Ia, where the observed optical bands probe the
rest-frame UV.  Establishing the luminosity distances of SNe~Ia is
extremely important for many studies, including the determination of
cosmological parameters such as $H_{0}, \Omega_{M}$, and
$\Omega_{\Lambda}$ \citep[e.g.,][]{Riess98:Lambda, Perlmutter99,
Riess05}.  Despite a large intrinsic dispersion of SN~Ia absolute
magnitudes ($\sigma \approx 0.7$~mag), there are ways of
``correcting'' this. For SNe~Ia with precise and independent relative
distance measurements, such as SNe~Ia in the Hubble flow or those in
host galaxies which have distances measured by the Cepheid
period-luminosity ($P/L$) relationship, one may test whether intrinsic
SN features correlate with peak luminosity.  Doing this, various authors
have found several observables in SN light curves and spectra that
correlate with luminosity \citep[e.g.,][]{Nugent95, Benetti05,
Wang07}.  The most well-known relationship is that between the
light-curve shape and luminosity \citep{Phillips93}.  We can then use
such relationships to determine the precise distances of high-redshift SNe.

Although there are many different known observables that correlate
with luminosity, thus far a combination has failed to significantly
reduce the scatter in Hubble diagrams, indicating that all known
luminosity indicators probe essentially the same parameter space,
precluding further improvements in SN~Ia distances.  Since the
luminosity of SNe~Ia (at peak) is powered by the decay of $^{56}$Ni
and UV spectra of SNe~Ia are highly dependent on iron-peak elements,
one might expect there to be correlations between observables in the
UV spectrum of a SN~Ia and its peak luminosity.

An indication that the UV colors might probe a different part of
parameter space than the light-curve shape vs.\ luminosity
relationship is found in the case of SNe~1992A and 1994D.  Despite
having very similar light-curve shapes (SN~1992A has $\Delta m_{15}
(B) = 1.33$ and SN~1994D has $\Delta m_{15} (B) = 1.31$; N.\ Suntzeff,
2007, private communication; \citealt{Richmond95}) and $B - V$ colors,
SN~1994D was \about 0.3~mag bluer than SN~1992A in the near-UV (as
measured by $U - B$).  This shows that there are real differences
among SNe~Ia that do not correlate with light-curve shape and suggests
that there could be a ``second parameter'' that may further reduce the
scatter in the  derived SN~Ia luminosities.

With this in mind, we have examined all 64 SN~Ia UV spectra observed
with either the {\it International Ultraviolet Explorer} ({\it IUE})
or {\it HST} through 2004, when STIS became inoperable.  For
completeness, we also present the additional 12 SN~Ia UV spectra 
having no available light-curve information.  Using our sample of 
SN~Ia spectra, we investigate possible correlations of the observables 
in the UV spectra with luminosity.

%%%%%%%%%%%%%%%%%%%%
%%  Observations  %%
%%%%%%%%%%%%%%%%%%%%

\section{Observations}

A total of 77 UV spectra of 20 SNe~Ia were obtained with {\it
IUE} and {\it HST} through 2004.  Of these, there are 16 SNe~Ia with
available optical light curves.  Most spectra were taken with {\it IUE},
with later spectra from {\it HST} using both the Faint Object
Spectrograph (FOS) and STIS.  Some of these data have been published
elsewhere (SN 1981B, \citealt{Branch86}; SNe 1990N, 1991T,
\citealt{Jeffery92}; SN 1992A, \citealt{Kirshner93}; SNe 2001eh, 2001ep,
\citealt{Sauer08}).  All data were retrieved from the {\it IUE} and
{\it HST} data archives \citep{deLaPena94}.  The fully reduced data
were then reprocessed to remove cosmic rays and other defects
according to standard published techniques.

The photometric and host-galaxy information for each SN is listed in
Table~\ref{t:phot}.  When available, we have included the distance
modulus for the host galaxies.  Distances are derived from infrared (IR) 
surface brightness fluctuations (SBF), Cepheids, or directly for hosts in 
the Hubble flow.  All distance moduli are on the IR SBF scale determined
by \citet{Jensen03}, calibrated to the {\it HST} Key Project Cepheid scale
\citep{Freedman01}, yielding $H_{0} = 76$~km~s$^{-1}$~Mpc$^{-1}$.

Having the distance modulus ($\mu$) for each host galaxy, the visual 
maximum brightness and measured host-galaxy visual extinction (from
MLCS2k2 \citealt{Jha07}) allow one to calculate the absolute
brightness of each SN at maximum light:
\begin{equation}
  M_{V} = V_{\text{max}} - A_{V} - \mu.
\end{equation}
To compare with the light-curve shape vs. luminosity relationship, we
have also included in Table~\ref{t:phot} two measures of light-curve
shape: $\Delta m_{15}$ \citep{Phillips93} and MLCS $\Delta$
\citep{Riess96, Jha07}.

There exists a single UV spectrum of the fast-declining, subluminous
SN~1991bg \citep[e.g.,][]{Filippenko92:91bg}; however, it has such a
low signal-to-noise ratio (S/N) that no clear SN signal can be
discerned.  It has been removed from all subsequent discussion, but is
included in Tables~\ref{t:phot} and \ref{t:spec} for completeness.
In addition to the 15 SNe~Ia (removing SN~1991bg) with available light
curves, there exist four objects (SNe~1989M, 1990M, 2003bf, and
2003bt) which have UV spectra, but no available light curves.  Without
light curves, the phase relative to maximum brightness, light-curve
shape parameters, and brightness at maximum cannot be determined; thus,
these objects have been removed from further analysis.  We do include
these SNe in Table~\ref{t:phot} and their spectra in Table~\ref{t:spec}
for completeness and for future studies if their light curves
eventually become available.

\begin{figure}
\epsscale{2.1}
\rotatebox{90}{
\plotone{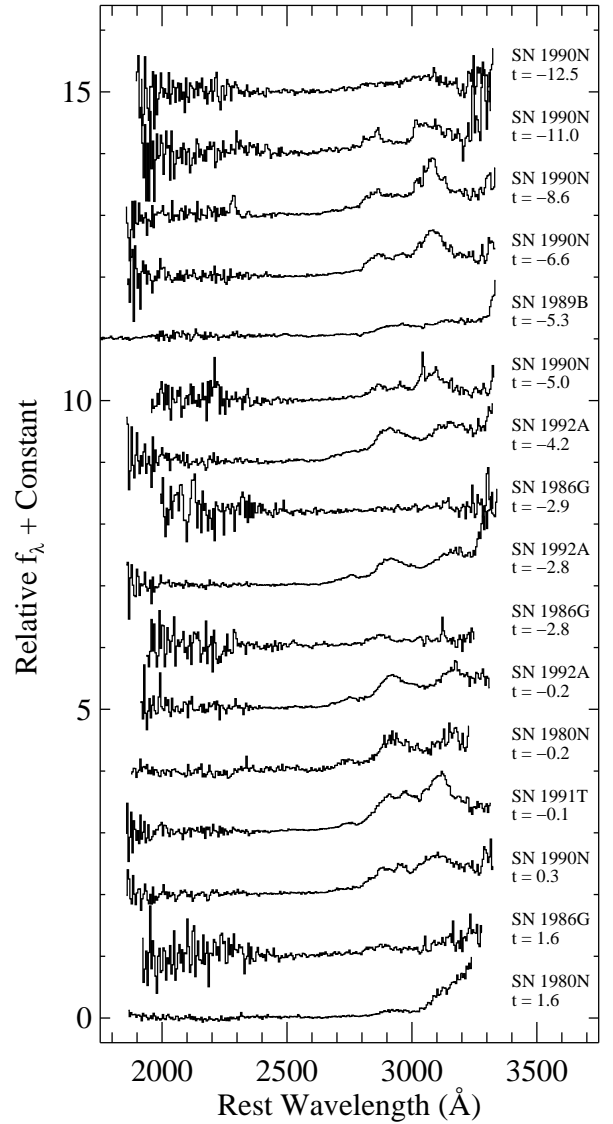}}
\caption{UV spectra of SNe~Ia with $t \leq 1.6$~d past maximum
brightness.  The spectra are generally of low S/N, but some features
are apparent.  In particular, there is an absorption trough at \about
3000~\AA, attributed to \ion{Fe}{2} $\lambda$3250 \citep{Branch86}, on
either side of two peaks.  These features are seen in nearly every
spectrum with $t < 3$~weeks and of reasonable S/N.}\label{f:spec1}
\end{figure}

\begin{figure}
\epsscale{2.1}
\rotatebox{90}{
\plotone{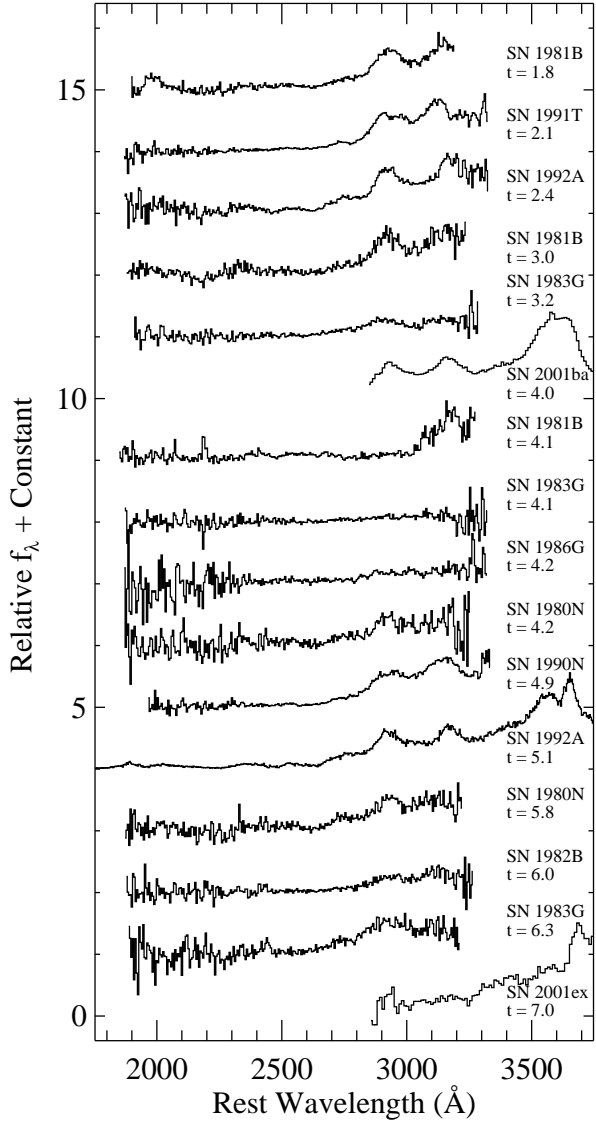}}
\caption{Same as Figure~\ref{f:spec1}, except with $1.8 \leq t \leq
7.0$~d past maximum.}\label{f:spec2}
\end{figure}

\begin{figure}
\epsscale{2.1}
\rotatebox{90}{
\plotone{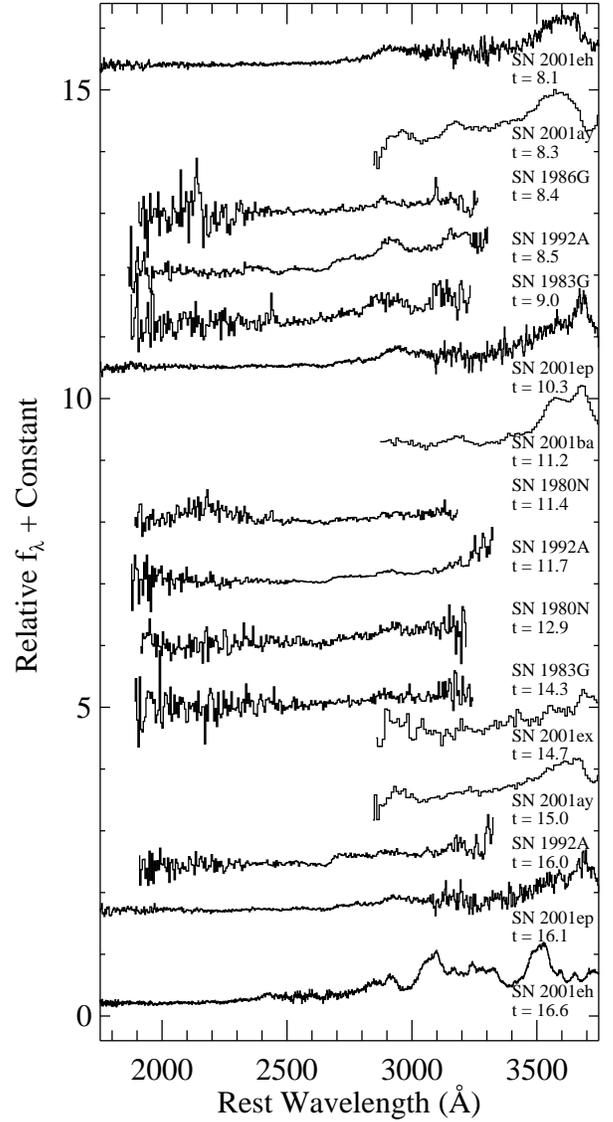}}
\caption{Same as Figure~\ref{f:spec1}, except with $8.1 \leq t \leq
16.6$~d past maximum.}\label{f:spec3}
\end{figure}

\begin{figure}
\epsscale{2.1}
\rotatebox{90}{
\plotone{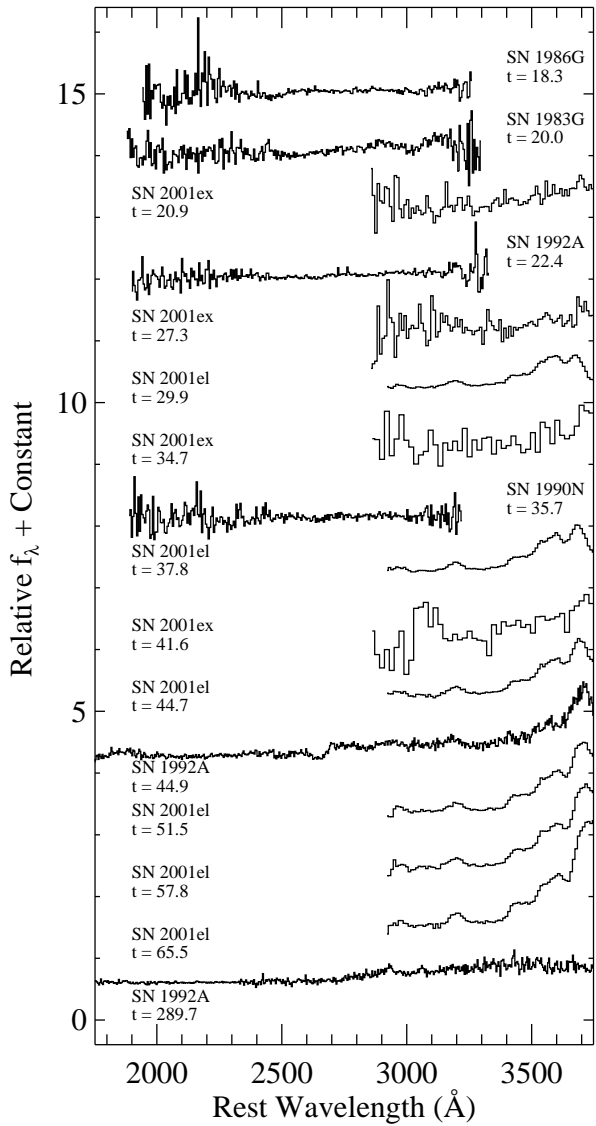}}
\caption{Same as Figure~\ref{f:spec1}, except with $t \geq
18.3$~d past maximum.}\label{f:spec4}
\end{figure}

\begin{figure}
\epsscale{1.7}
\rotatebox{90}{
\plotone{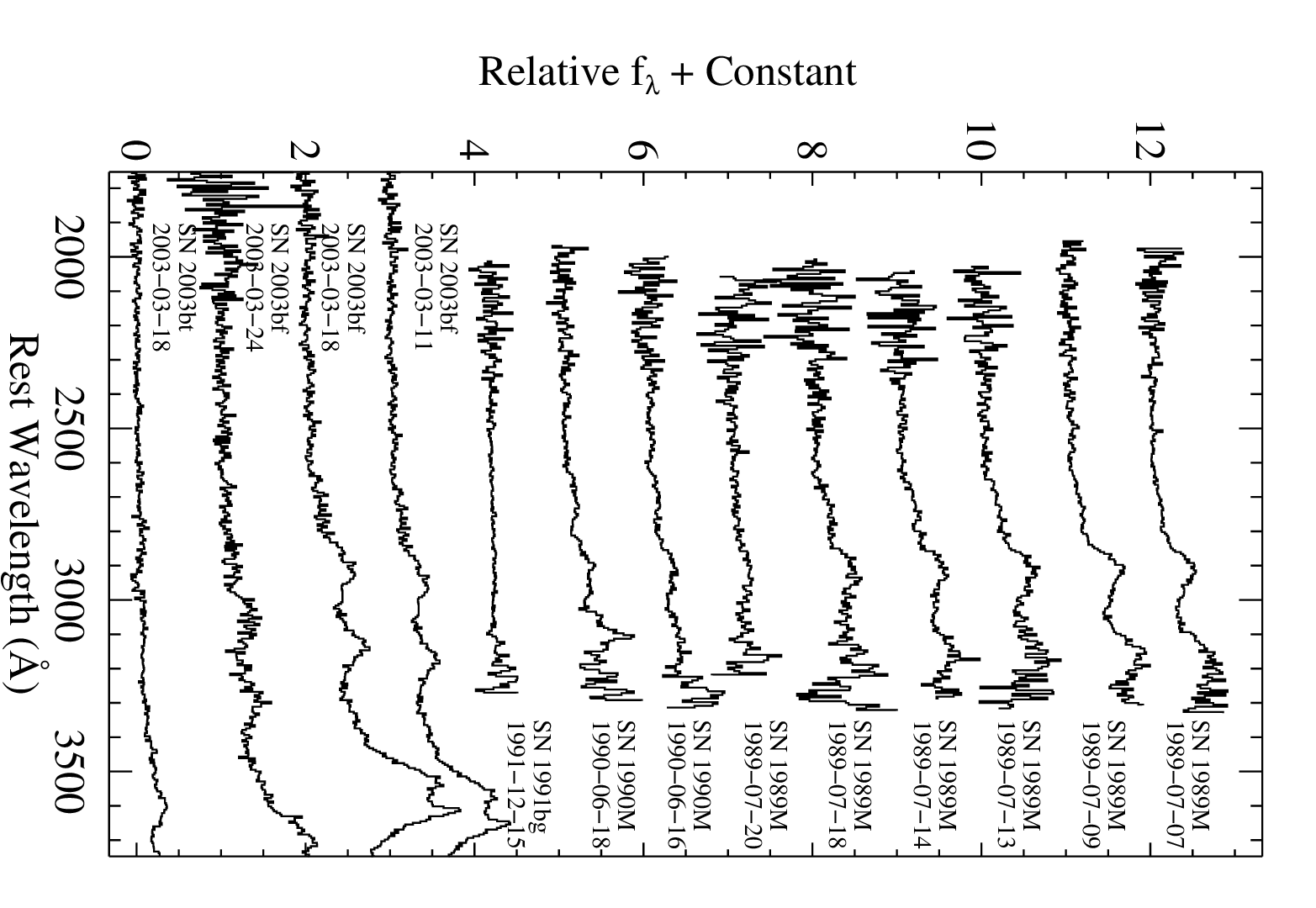}}
\caption{UV spectra of SNe~Ia with no phase information. UT dates
of observation are given.  We also include a spectrum of SN~1991bg,
taken at $t = 2.4$~d, but with no discernible signal.}\label{f:spec5}
\end{figure}

We also provide specific information for each spectrum in
Table~\ref{t:spec}, listing the UT date of observation, the telescope
(and instrument) used to obtain the spectrum, the phase of the
spectrum relative to maximum $B$ brightness (when available;
determined from the light curves; see Table~\ref{t:phot}), and various
quantities calculated from the spectra (see \S~\ref{s:analysis}).  The
64 spectra with phase information are presented in
Figures~\ref{f:spec1}--\ref{f:spec4}.  The 13 spectra of SNe~1989M,
1990M, 1991bg, 2003bf, and 2003bt (which have no phase information)
are presented in Figure~\ref{f:spec5}.  All spectra compiled in this paper
can be obtained
online\footnote{http://astro.berkeley.edu/$\sim$rfoley/uvspectra/ .}.  For
all analyses, the spectra have been corrected for Galactic extinction
\citep{Schlegel98}, and $R_{V} = 3.1$ has been assumed for both Milky
Way and host-galaxy extinction.

%%%%%%%%%%%%%%%%
%%  Analysis  %%
%%%%%%%%%%%%%%%%

\section{Analysis}\label{s:analysis}

\subsection{Line Features}

Although the optical maximum-light spectra of SNe~Ia are dominated
mainly by intermediate-mass elements \citep[e.g., Ca, Si, S; see][for
a review]{Filippenko97}, the UV maximum-light spectrum is dominated by
Fe-peak elements.  An analysis of the $t = 5$~d spectrum of SN~1992A
has revealed many complex, low-amplitude line features
\citep{Kirshner93}, but for almost all of our sample the S/N is too
low to detect them. We do, however, detect the prominent trough at
\about3000~\AA, attributed to \ion{Fe}{2} $\lambda$3250
\citep{Branch86}, in nearly all of our spectra having sufficiently
high S/N at $t < 3$~weeks (Fig.~\ref{f:spec1}--\ref{f:spec4}). Being
the strongest feature in the near-UV spectra of SNe~Ia, we are able to
measure its velocity and equivalent width (EW) in most of our data.

\begin{figure}
\epsscale{0.85}
\rotatebox{90}{
\plotone{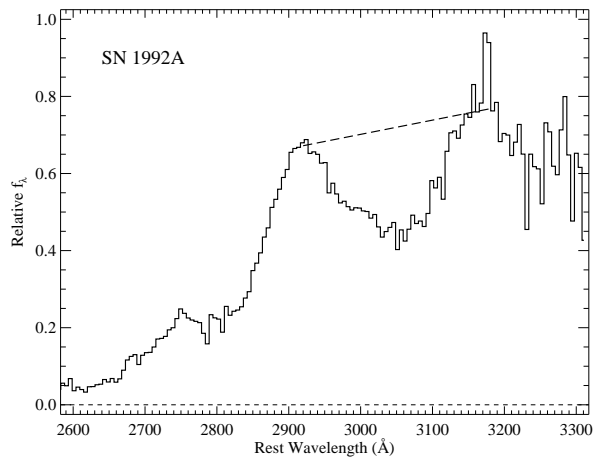}}
\caption{UV spectrum of SN~1992A at $t = -0.2$~d.  An example of the
choice of the continuum placement for measuring the EW of \ion{Fe}{2}
$\lambda 3250$ is shown.}\label{f:spec_ew}
\end{figure}

We fit a Gaussian profile to the \ion{Fe}{2} feature to measure its
velocity.  Since the S/N of the spectra is relatively low, we also
use the Gaussian fits to measure the EW of the \ion{Fe}{2} line.  To
properly measure the \ion{Fe}{2} EWs, we must choose wavelength
regions to act as the continuum.  For this, we select the median value
from the peaks just blueward and redward of the line (an example can
be seen in Figure~\ref{f:spec_ew}).  These regions are different for
each spectrum and are somewhat subjective.  Consequently, the
systematic errors associated with any given measurement may be as
large as 10~\AA.

\begin{figure}
\epsscale{0.85}
\rotatebox{90}{
\plotone{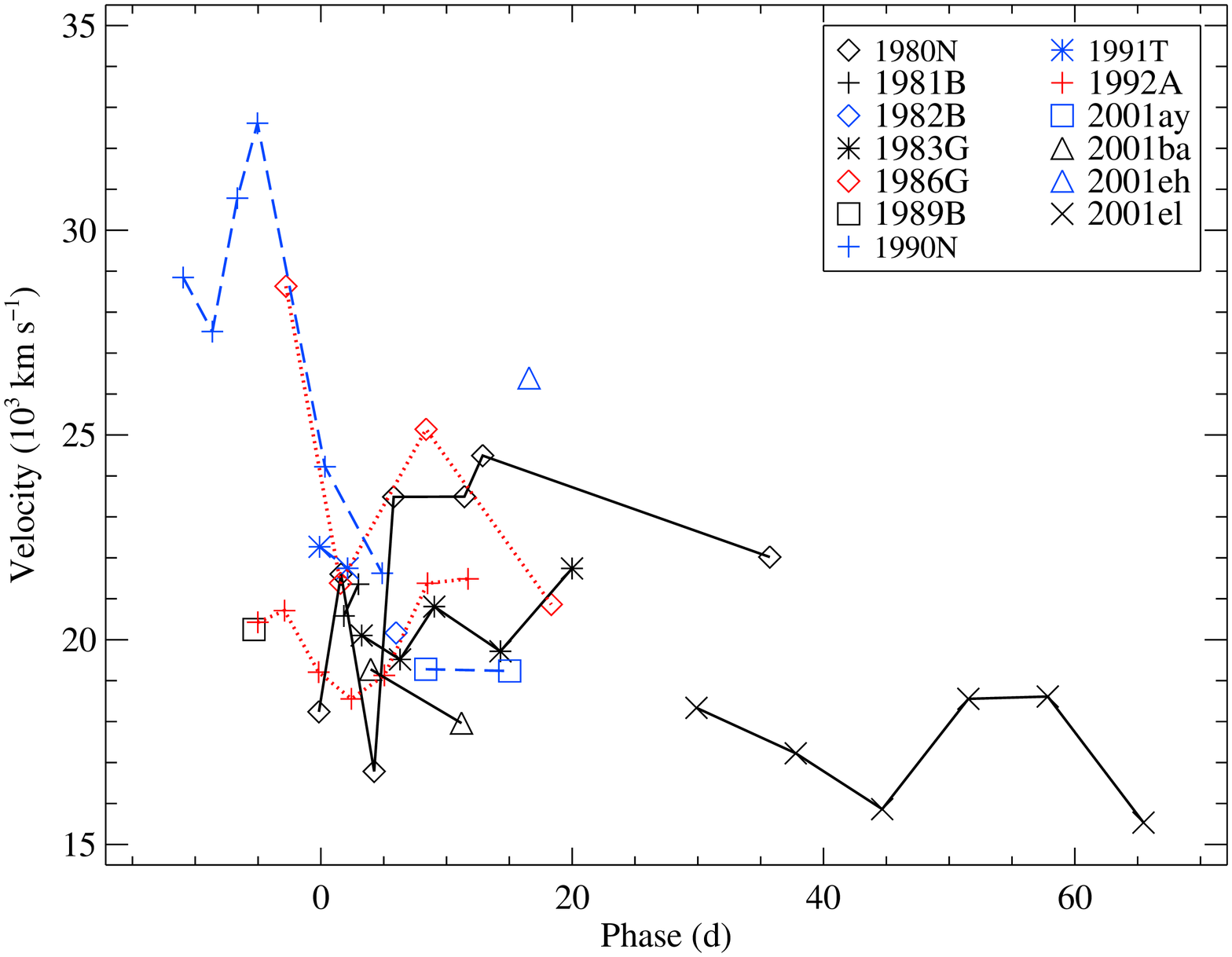}}
\caption{The blueshifted velocity of the minimum of the \ion{Fe}{2}
feature as a function of time for various SNe~Ia assuming a rest-frame
$gf$-weighted wavelength of 3250~\AA\ for \ion{Fe}{2}.  The normal,
high, and low-$\Delta$ (corresponding to normal, low, and high
luminosity) objects are shown in black, red, and blue (with solid,
dotted, and dashed lines connecting points),
respectively.}\label{f:vel_age}
\end{figure}

In Figure~\ref{f:vel_age}, we present the \ion{Fe}{2} line velocity as
a function of time for our sample.  We assume that the \ion{Fe}{2}
feature has a $gf$-weighted rest-frame wavelength of 3250~\AA\
(D. Branch, 2007, private communication).  Although the exact
rest-frame wavelength of the line may be slightly different from this
value, it is not important for examining general trends and comparing
objects.  For the \ion{Fe}{2} feature, we see that the velocity
decreases significantly from premaximum to maximum light for SNe~1986G
and 1990N, while the velocity change is relatively flat for
SNe~1992A.  The velocity gradient is relatively shallow after maximum
brightness.  With few pre-maximum spectra, we are unable to
definitively say whether the velocity gradient for typical SNe~Ia
decreases with time similar to the velocity evolution of optical lines
from intermediate-mass elements \citep[e.g.,][]{Patat96, Benetti05}
or if the velocity gradient is flatter.  We do not see any trends in
the velocity evolution that correlate with light-curve shape.

\begin{figure}
\epsscale{0.85}
\rotatebox{90}{
\plotone{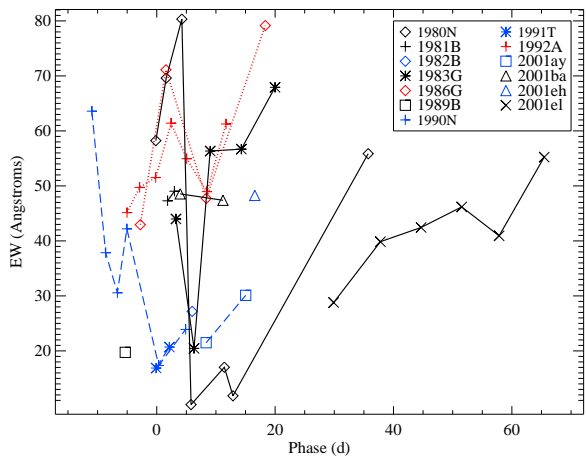}}
\caption{The EW of the \ion{Fe}{2} feature as a function of time for
various SNe~Ia.  The normal, high, and low-$\Delta$ (corresponding to
normal, low, and high luminosity) objects are shown in black, red, and
blue (with solid, dotted, and dashed lines connecting points),
respectively.}\label{f:ew_age}
\end{figure}

As seen in Figure~\ref{f:ew_age}, the \ion{Fe}{2} EWs are
significantly different from object to object.  Moreover, there is no
definitive trend among the entire sample for the evolution of the line
with time.  For example, the EW in SN~1980N increases  from around
maximum brightness for about a week, then falls dramatically (87\%
over 1.6~d), only to climb to its previous level about a month after
maximum brightness.

However, separating the objects by light-curve shape does reveal some
differences among the subsamples.  On average, the low-$\Delta$
(slow-declining light curve, high-luminosity) objects have lower EWs
near maximum, and for the single object with several pre-maximum
epochs, the EW appears to decline with time before maximum brightness.
The high-$\Delta$ (fast-declining light curve, low-luminosity)
objects, on the other hand, have larger EWs near maximum brightness,
and the strength of the line appears to increase with time from before
maximum until after maximum.  The normal-luminosity events do not
appear to follow any particular trend, but their EWs tend to fall
between the high and low-$\Delta$ objects at maximum brightness.

\subsection{UV Spectral Shape and Absolute Magnitude}\label{ss:spec_shape}

\begin{figure}
\epsscale{0.85}
\rotatebox{90}{
\plotone{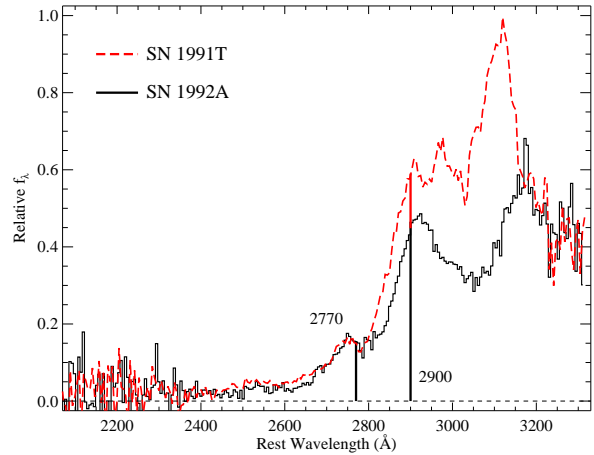}}
\caption{UV spectrum of SN~1992A at $t = -0.2$~d (black) and SN~1991T
at $t = -0.1$~d (red) normalized to have the same flux at 2770~\AA.
The fluxes measured at 2770 and 2900~\AA\ are shown.  SN~1991T has a
smaller UV ratio (see Eq.~\ref{e:ruv}) than
SN~1992A.}\label{f:spec_ratio}
\end{figure}

Examining the UV spectra, we see that the overall spectral shape
differs dramatically among various objects.  Some objects have strong
continua at \about 3200~\AA, only to drop rapidly by \about
2600~\AA, while others have a much shallower decline from long to
short wavelengths.  Figure~\ref{f:spec_ratio} shows spectra of
SNe~1991T and 1992A at maximum light.  It is clear that SN~1991T has
excess flux relative to SN~1992A in the range $2900 \lesssim \lambda
\lesssim 3100$~\AA.

Although there are many ways to characterize the spectral slope (such
as fitting a straight line to the flux density), we determined that
differences in spectral features prevent an accurate assessment of
the slope with most of these methods.  One method which can avoid
spectral features (to a large degree) is to measure the ratio of the
flux density at two separate wavelengths.  For this, we have chosen
wavelengths of 2770 and 2900~\AA, which correspond to parts of the
spectra that have reasonable S/N and avoid major features in the
near-maximum ($-3 \le t \le 3$~d) spectra.  We define the ``UV ratio''
as
\begin{equation}
  \mathcal{R}_{UV} = f_{\lambda} (2770~\text{\AA}) / f_{\lambda}
    (2900~\text{\AA}).
\end{equation}
For $f_{\lambda} (2770~\text{\AA})$ and $f_{\lambda}
(2900~\text{\AA})$, we measure the median flux density in the 40~\AA\
regions centered on 2770 and 2900~\AA, respectively.
Figure~\ref{f:spec_ratio} demonstrates this method with SNe~1991T and
1992A. We also calculate an uncertainty for this ratio by measuring
all flux-density ratios within a 20~\AA\ region centered on each
wavelength value ($\pm 10$~\AA) and with the same wavelength spacing
to achieve a corresponding slope [e.g., $f_{\lambda}
(2760~\text{\AA}) / f_{\lambda} (2890~\text{\AA})$, $f_{\lambda}
(2765~\text{\AA}) / f_{\lambda} (2895~\text{\AA})$, $f_{\lambda}
(2780~\text{\AA}) / f_{\lambda} (2910~\text{\AA})$].  The UV ratio and
its uncertainty are presented in Table~\ref{t:spec}  for all objects
covering the wavelength range 2750--2920~\AA.

We chose the range $-3 < t < 3$~d so that the features do not shift
dramatically from one spectrum to another causing the adopted fixed
wavelengths to correspond to different physical features.  If we
expand the range of phases to $\pm 7$~d, the relationships presented
below remain, but with increased scatter.

\begin{figure}
\epsscale{0.85}
\rotatebox{90}{
\plotone{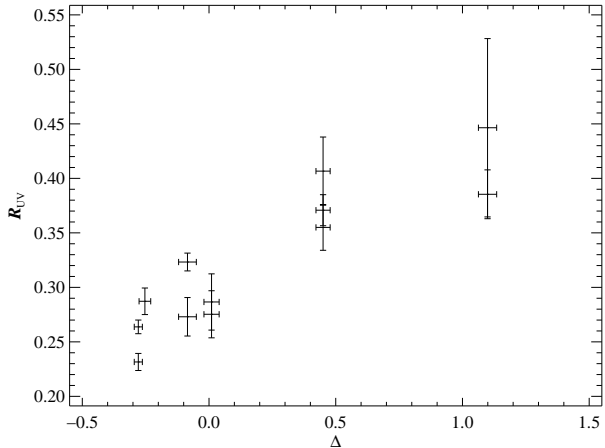}}
\caption{Relationship between the UV ratio and $\Delta$ for spectra
with $-3 < t < 3$d.}\label{f:delta_ratio}
\end{figure}

In Figure~\ref{f:delta_ratio}, we see that the UV ratio and $\Delta$
are highly correlated (correlation coefficient of 0.900).
Objects that decline slowly (small $\Delta$) have small UV ratios,
while those that fade rapidly (large $\Delta$) have large UV ratios.
This is similar to stating that SNe~Ia with slow-declining light 
curves have redder spectra between 2770 and 2900~\AA\ than those with
fast-declining light curves.  This is opposite to the trend in optical 
colors, where SNe~Ia with slow-declining light curves have bluer $U-B$ 
and $B-V$ than those with fast-declining light curves.

\begin{figure}
\epsscale{0.85}
\rotatebox{90}{
\plotone{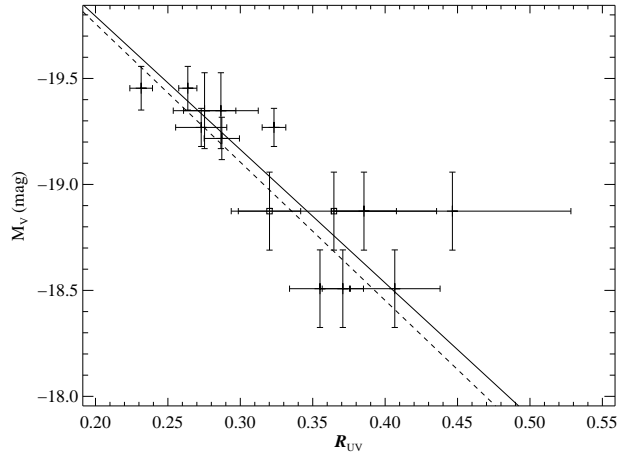}}
\caption{The UV ratio compared to the absolute magnitude at
maximum for several SNe~Ia.  The square points are the values for
SN~1986G if no host-galaxy reddening correction is applied to the
spectrum before calculating $\mathcal{R}_{UV}$ (an extinction
correction is still applied to its measured absolute magnitude).  The
solid line is the fit for the host-galaxy extinction corrected
spectra, described by $M_{V} = -19.165 + 6.293 (\mathcal{R}_{UV} -
0.3)$~mag, where $\mathcal{R}_{UV}$ is the UV flux ratio.  The dashed
line is the fit for the uncorrected spectra, described by $M_{V} =
-19.105 + 6.527 (\mathcal{R}_{UV} - 0.3)$~mag.}\label{f:ratio_mv}
\end{figure}

In Figure~\ref{f:ratio_mv}, we show the UV ratio of 12 spectra (from 6
objects: SNe~1980N, 1981B, 1986G, 1990N, 1991T, and 1992A) with $-3
\le t \le 3$~d compared to $M_{V}$ at maximum light for the SNe. Spectra
of more recent SNe~Ia tend to have $t > 3$~d, and several barely probe the
UV, not reaching 2770~\AA.  We find that the UV ratio is highly
correlated with the peak absolute magnitude of the SN, with
correlation coefficients of 0.811 and 0.860 for the individual UV
ratio measurements for each spectrum and the averaged UV ratio
measurements for each object, respectively.  By averaging the UV ratio
for each object (to weight all objects equally), we have fit a line to
the data, yielding the relationship
\begin{equation}\label{e:ruv}
  M_{V} (t = 0) = -19.165 + 6.293 (\mathcal{R}_{UV} - 0.3) \text{ mag}
\end{equation}
with a residual scatter of 0.207~mag.  

\begin{figure}
\epsscale{0.85}
\rotatebox{90}{
\plotone{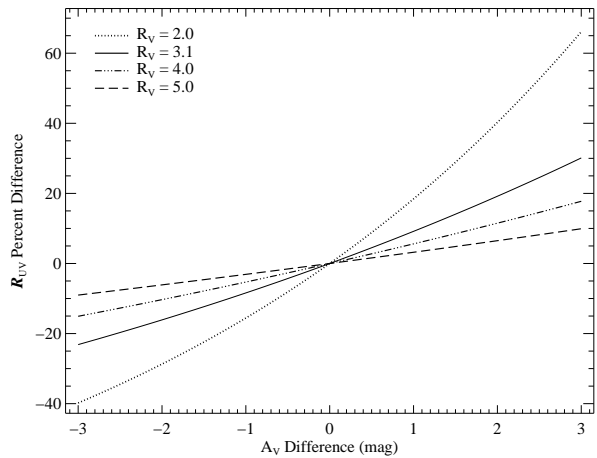}}
\caption{The percent difference of the UV ratio as a function of
$A_{V}$ for different values of $R_{V}$.  Negative values of $A_{V}$
are shown in case the extinction is overestimated.  For $R_{V} = 3.1$
($R_{V} = 2.0$), there is less than a 10\% difference for $-1.20 <
A_{V} < 1.08$~mag ($-0.66 < A_{V} < 0.60$).}\label{f:ratio_av}
\end{figure}

The measurement of $\mathcal{R}_{UV}$ is affected by dust reddening
along the line of sight (both from our Galaxy and the SN host galaxy).
Though this can be significant in the near-UV, the small wavelength
range over which we measure $\mathcal{R}_{UV}$ (only 130~\AA) makes it
rather insensitive to extinction. In Figure~\ref{f:ratio_av}, we show
the effect of dust reddening on the measurement of $\mathcal{R}_{UV}$,
and find that a $\pm 0.6$~mag change in $A_{V}$ yields less than a
10\% change in $\mathcal{R}_{UV}$ (for any $R_{V} \ge 2.0$).  Of the
sample presented here, only SN~1986G, which has $A_{V} = 2.313$~mag
(as measured from MLCS, assuming $R_{V} = 3.1$; for lower values of
$R_{V}$, $A_{V}$ also decreases), has a UV ratio significantly
affected by extinction.  If we were to ignore this effect, we would
have derived
\begin{equation}
  M_{V} (t = 0) = -19.105 + 6.527 (\mathcal{R}_{UV( \text{Uncor})} -
    0.3) \text{ mag},
\end{equation}
differing little from the relation in Eq.~\ref{e:ruv}.

\subsection{Reducing the Scatter in the SN~Ia Hubble Diagram}

As mentioned previously, combining various luminosity indicators has
failed to reduce the scatter in SN~Ia Hubble diagrams.  Therefore, a
single parameter can characterize the luminosity of a SN~Ia.

\citet{Jha07} derived $M_{V}$ as a quadratic function of $\Delta$.
Since our sample is different from the Hubble-flow sample used to
create the relationship, and the relationship of \citet{Jha07} is nearly
linear, and we only have 6 objects with measured $\mathcal{R}_{UV}$
and $-3 \leq t \leq 3$~d, we decided to fit for a new, linear relationship.
Excluding SN~1986G from our fits based on its large and uncertain
host-galaxy reddening, we find
\begin{equation}\label{e:nouv}
  M_{V}(t = 0) = -19.118 + 1.020 \Delta \text{ mag}.
\end{equation}
This relationship has $\chi^{2}$ per degree of freedom (dof) =
5.771/3, a reasonable, but not excellent, fit.  Using the absolute
magnitudes from this new relationship, we find a scatter in the
residuals to the absolute-magnitude fit of 0.167~mag, which is
slightly lower than, but similar to 0.18~mag, the scatter in the
Hubble diagram using the results of \citet{Jha07}.

If we then include the UV ratio, $\mathcal{R}_{UV}$, in the fit, we
find
\begin{align}
  M_{V}(t = 0) &= -19.161 + 0.043 \Delta \notag \\
    &+ 6.578 (\mathcal{R}_{UV} - 0.3) \text{ mag}.\label{e:uv}
\end{align}
This relationship has $\chi^{2} / {\rm dof} = 2.155 / 2$, which is
much improved from the relationship given in Eq.~\ref{e:nouv}.  The
residual scatter also decreases to 0.090~mag.  The low scatter (the
mean error for the absolute magnitudes is 0.140~mag) indicates that
the data may be slightly overfit, requiring more data to determine if
this relationship is valid.  Eq.~\ref{e:uv} shows that
$\mathcal{R}_{UV}$ is more influential on the determination of $M_{V}$
than $\Delta$.  In fact, fitting a relationship between just $M_{V}$
and $\mathcal{R}_{UV}$ (and removing SN~1986G from the sample) gives
$\chi^{2} / {\rm dof} = 2.166 / 3$.  It is obvious that we need a
larger sample to determine if $\mathcal{R}_{UV}$ is in the same family
of luminosity indicators as light-curve shape.

\begin{figure}
\epsscale{1.6}
\rotatebox{90}{
\plotone{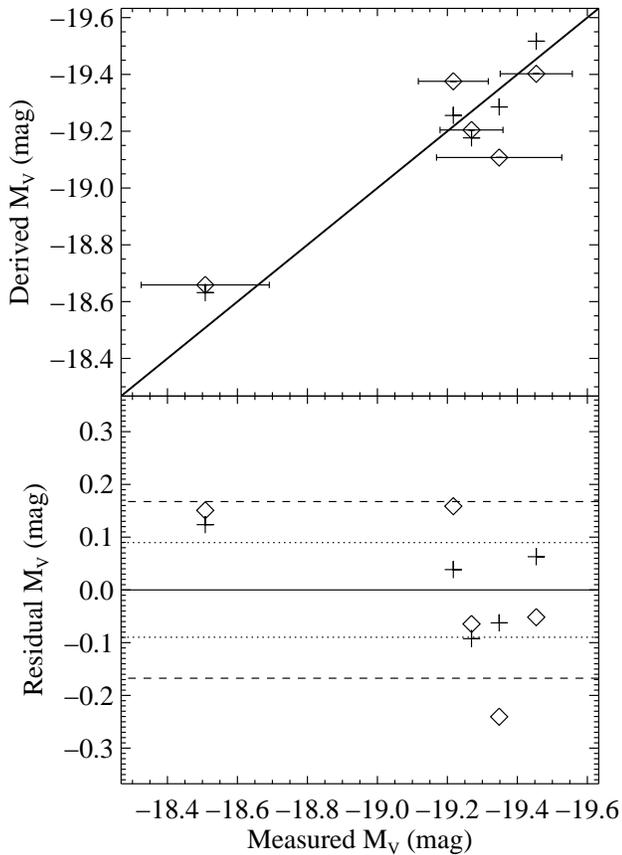}}
\caption{({\it top panel}):  Derived absolute magnitudes using
Eq.~\ref{e:nouv} (diamonds) and Eq.~\ref{e:uv} (crosses), which do not
and do include the UV ratio, respectively.  The uncertainties in the
measured $M_{V}$ values are shown.  ({\it bottom panel}):  The
residual absolute magnitudes from the relationships.  The dashed and
dotted lines show the standard deviation of the residuals from
Eq~\ref{e:nouv} and Eq.~\ref{e:uv}, respectively.}\label{f:abs_imp}
\end{figure}

Figure~\ref{f:abs_imp} shows the derived absolute magnitudes (and
residuals) from Eqs.~\ref{e:nouv} and \ref{e:uv} for our sample
of objects with near-maximum spectra.

%%%%%%%%%%%%%%%%%%
%%  Discussion  %%
%%%%%%%%%%%%%%%%%%

\section{Discussion and Conclusions}

\subsection{The \ion{Fe}{2} Feature}

Based on the very small sample of SNe~Ia with available UV spectra and
optical light curves, we are able
to see general trends regarding the evolution of the strong
\ion{Fe}{2} $\lambda$3250 feature.  The line velocity and its
evolution do not appear to differ significantly among objects with
different light-curve shapes. On the other hand, the strength near 
maximum light and evolution of the EW of the line do appear to
depend on the subsample, with fast-declining objects having a larger EW
near maximum light than slow-declining objects.

The variation in EW measurements for these subsamples is similar to
the difference in the EW of the \ion{Si}{2} $\lambda 5972$ line seen
in SN~1991T-like and SN~1991bg-like objects
\citep[e.g.,][]{Filippenko92:91bg, Filippenko92:91T}, and may be
another indication of luminosity.

The trend between the slope of the UV continuum (i.e., the UV ratio)
and the absolute magnitude of the SN, with brighter objects having a
steeper decline to shorter wavelengths, is similar to stating that
more-luminous objects have brighter near-UV flux.  The difference
between \ion{Fe}{2} EWs for the slow and fast-declining objects may
simply be a higher continuum in the slow-declining objects.
Similarly, SN~1991T and its brethren have hot photospheres which
ionize Fe to \ion{Fe}{3}, and at maximum light they have strong
\ion{Fe}{3} features and absent \ion{Fe}{2} features, which are
present in normal-luminosity SNe~Ia \citep[e.g.,][]{Filippenko92:91T}.
The \ion{Fe}{2} $\lambda$3250 feature may be weaker in the UV spectra
of SN~1991T-like objects because most of  the Fe has been ionized to
\ion{Fe}{3}.  Since the UV continuum is blanketed by \ion{Fe}{2}
lines, the \ion{Fe}{2}/\ion{Fe}{3} ratio may be the reason for the
differences in both the \ion{Fe}{2} EW and UV ratio.

\subsection{The UV Continuum}

We have shown that for our small sample, there is a relationship
between the near-UV spectral slope, described by our ``UV ratio,'' and
the luminosity of SNe~Ia.  For a given SN~Ia, the luminosity is
derived from a combination of distance moduli from independent
sources, a measured peak apparent magnitude, and a calculated
extinction.  The UV slope is measured from the spectrum of a SN.
Since these two observations are independent, it is very intriguing
that we are able to find a strong correlation between them.  This new
luminosity indicator might turn out to be particularly useful for high-redshift
objects where it is difficult to observe the rest-frame optical light.

We show that slow-declining SNe~Ia have redder near-UV colors (as
measured by the UV ratio) than fast-declining SNe~Ia.  \citet{Jha06}
found that slow-declining SNe~Ia have bluer $U-B$ colors than fast
declining SNe~Ia.  This difference suggests that the near UV, in
particular \about2800--4000~\AA, may be very sensitive to luminosity,
with bright, slow-declining SNe~Ia having a larger flux density in
this region compared to faster-declining objects.  This would lead to
both redder $\mathcal{R}_{UV}$ values and bluer $U - B$ colors.
Combining this result with recent studies of the variation in SN~Ia
ground-based near-UV spectra \citep{Foley08, Ellis08} may suggest that
the region strongly correlated with luminosity might be restricted to
the range \about 2800--3500~\AA.

We also present the possibility that the relationship between the UV
ratio and luminosity may represent a new vector in the
multi-dimensional SN~Ia luminosity parameter space.  This possible
``second parameter'' might open the door for more precise distances to
high-redshift SNe~Ia, significantly improving our measurements of
cosmological parameters, such as the equation-of-state parameter of
the dark energy, $w = P/(\rho c^2)$, and its derivatives.

\subsection{Future Studies}

One way to test the validity of the UV ratio as a luminosity indicator
is to measure a similar photometric color.  The wavelengths we chose
for this study, 2770 and 2900~\AA, are very closely spaced, and it is
difficult to probe flux densities at these wavelengths independently.
However, the low S/N of our spectra prevented us from probing wavelengths 
shortward of 2600~\AA, and the UV spectra rarely included wavelengths 
redward of 3300~\AA.  Use of longer or shorter wavelengths may 
provide a similar (or potentially better) relationship.  However,
widely separated wavelengths make the results more sensitive to errors
in the adopted reddening.

There are currently several SNe~Ia that have been observed
photometrically in the UV with the \emph{UVOT} instrument on
\emph{Swift}.  Its UVW2, UVM2, and UVW1 filters have effective
wavelengths of 1900, 2200, and 2600~\AA, respectively.  Either a
$\text{UVM2} - \text{UVW1}$ or $\text{UVW1} - U$ color may result in a
similar relationship.  We do note, however, that the UVW1 and UVW2
filters have long red tails in their response functions; combined 
with the rapidly increasing flux of a SN~Ia to the red, this may
increase the observed scatter in the relationship.

Additional high-S/N spectra of SNe~Ia with {\it HST} or {\it Swift}
(which can obtain very early epochs) in host galaxies having precise
relative distances are also necessary to test this relationship.  The
small sample of 6 SNe~Ia with available maximum-light UV spectra and
optical light curves provides a glimpse of the potential UV spectra
have in determining SN~Ia distances.  It may be that this relationship
is a statistical fluke; however, the light-curve shape vs. luminosity
relationship was first clearly measured with a sample of only 7
objects (removing SNe~1986G and 1989B from a set of 9 because of their
uncertain reddening; \citealt{Phillips93}).  This gives us hope
that with a larger sample, the relationship will improve and provide
even better SN~Ia distances. With the anticipated repair of STIS
aboard {\it HST}, or at least with the installation of the Cosmic
Origins Spectrograph (COS), soon it should again be possible to obtain
high-quality UV spectra of SNe~Ia.

Most past and present high-redshift SN surveys have obtained limited
and typically poor rest-frame UV spectra for high-redshift SNe~Ia.
Currently there is no public database of high-redshift SN~Ia spectra for
which we can apply this method and test its validity at high redshifts.
\citet{Ellis08} present a large sample of high-S/N SN~Ia spectra,
including 12 objects which have spectra at $-3 \le t \le 3$~d.  This, 
or a similar sample, would be excellent for testing this method.

Looking forward, the Joint Dark Energy Mission (JDEM) will find and
follow several thousand SNe~Ia out to $z \approx 2$.  Incorporating
rest-frame UV photometry and spectroscopy into both mission planning
and analysis may help produce the requisite precision necessary to achieve
its mission objectives.

\begin{acknowledgments}
We are grateful to D. Branch for revisiting research he performed over
20 years ago.  All of the data presented in this paper were obtained
from the Multimission Archive at the Space Telescope Science Institute
(MAST), including observations made with the NASA/ESA {\it Hubble
Space Telescope}.  STScI is operated by the Association of
Universities for Research in Astronomy, Inc., under NASA contract
NAS5--26555.  Support for MAST for non-{\it HST} data is provided by
the NASA Office of Space Science via grant NAG5--7584 and by other
grants and contracts.  Our work is supported by NSF grant
AST--0607485, as well as by NASA/{\it HST} grant GO--10182 from STScI.
This research has made extensive use of the NASA/IPAC Extragalactic
Database (NED), which is operated by the Jet Propulsion Laboratory,
California Institute of Technology, under contract with NASA.

{\it Facilities:} {\it IUE}, {\it HST}: FOS, STIS

\end{acknowledgments}

\bibliographystyle{apj}
\bibliography{astro_refs}

\begin{center}
\tabletypesize{\tiny}
\clearpage
\begin{deluxetable}{llclccccccc}
%\rotate
\tabletypesize{\tiny}
\tablewidth{0pt}
\tablecaption{Photometric and Host-Galaxy Information\label{t:phot}}
\tablehead{
\colhead{SN} &
\colhead{Host Galaxy} &
\colhead{$\mu$} &
\colhead{Method\tablenotemark{a}} &
\colhead{$\mu$} &
\colhead{Date of $B$ Maximum} &
\colhead{$V_{\text{max}}$} &
\colhead{$\Delta m_{15}(B)$} &
\colhead{$\Delta$} &
\colhead{$A_{V}$} &
\colhead{Phot} \\
\colhead{Name} &
\colhead{} &
\colhead{(mag)} &
\colhead{} &
\colhead{Ref} &
\colhead{(JD$-$2,400,000)} &
\colhead{(mag)} &
\colhead{} &
\colhead{} &
\colhead{(mag)} &
\colhead{Ref}}

\startdata

SN~1980N  & NGC~1316                & 31.50 (0.17) & SBF     & 1       & 44585.38 (0.35) & 12.459 (0.034) & 1.076 (0.089) & \phs0.010 (0.030) & 0.307 (0.044) & 2        \\
SN~1981B  & NGC~4536                & 30.87 (0.04) & C       & 3       & 44670.95 (0.43) & 11.954 (0.033) & 1.009 (0.088) &  $-0.085$ (0.035) & 0.353 (0.074) & 4,5      \\
SN~1982B  & NGC~2268                & \nodata      & \nodata & \nodata & 45012.91 (1.06) & 13.333 (0.046) & 0.940 (0.105) &  $-0.190$ (0.060) & 0.143 (0.125) & 6        \\
SN~1983G  & NGC~4753                & 31.70 (0.19) & SBF     & 7       & 45429.72 (0.68) & 12.669 (0.036) & 1.093 (0.092) & \phs0.063 (0.081) & 0.924 (0.103) & 8,9      \\
SN~1986G  & NGC~5128                & 27.96 (0.14) & SBF     & 7       & 46561.43 (0.13) & 11.399 (0.065) & 1.626 (0.135) & \phs1.099 (0.036) & 2.313 (0.101) & 10       \\
SN~1989B  & NGC~3627                & 30.01 (0.08) & C       & 3       & 47564.27 (0.41) & 11.993 (0.036) & 1.066 (0.091) & \phs0.046 (0.063) & 1.457 (0.070) & 11       \\
SN~1989M  & NGC~4579                & \nodata      & \nodata & \nodata & \nodata         & \nodata        & \nodata       & \nodata           & \nodata       & \nodata  \\
SN~1990M  & NGC~5493                & \nodata      & \nodata & \nodata & \nodata         & \nodata        & \nodata       & \nodata           & \nodata       & \nodata  \\
SN~1990N  & NGC~4639                & 31.71 (0.08) & C       & 3       & 48081.86 (0.08) & 12.738 (0.035) & 0.890 (0.091) &  $-0.253$ (0.023) & 0.245 (0.049) & 12       \\
SN~1991T  & NGC~4527                & 30.61 (0.09) & C       & 13      & 48374.16 (0.07) & 11.493 (0.034) & 0.871 (0.089) &  $-0.279$ (0.016) & 0.337 (0.037) & 14,15    \\
SN~1991bg & NGC~4374 (M~84)         & 31.16 (0.05) & SBF     & 7       & 48602.91 (0.20) & 11.913 (0.101) & 0.871 (0.089) & \phs1.409 (0.026) & 0.467 (0.065) & 16,17,18 \\
SN~1992A  & NGC~1380                & 31.07 (0.18) & SBF     & 1       & 48640.63 (0.11) & 12.578 (0.032) & 1.362 (0.088) & \phs0.450 (0.028) & 0.016 (0.011) & 15,19    \\
SN~2001ay & IC~4423                 & 35.57 (0.15) & HF      & 20      & 52023.71 (0.49) & 16.588 (0.034) & 0.759 (0.088) &  $-0.441$ (0.026) & 0.338 (0.055) & 21       \\
SN~2001ba & MCG~-05-28-001          & 35.55 (0.15) & HF      & 20      & 52034.21 (0.27) & 16.551 (0.046) & 1.014 (0.105) &  $-0.116$ (0.029) & 0.016 (0.012) & 22       \\
SN~2001eh & UGC~1162                & 35.92 (0.15) & HF      & 20      & 52169.56 (0.19) & 16.815 (0.045) & 0.830 (0.104) &  $-0.307$ (0.020) & 0.024 (0.016) & 23       \\
SN~2001el & NGC~1448                & \nodata      & \nodata & \nodata & 52182.13 (0.08) & 12.713 (0.033) & 0.995 (0.087) &  $-0.090$ (0.020) & 0.745 (0.040) & 24       \\
SN~2001ep & NGC~1699                & \nodata      & \nodata & \nodata & 52200.10 (0.08) & 14.989 (0.040) & 1.109 (0.097) & \phs0.057 (0.026) & 0.568 (0.056) & 23       \\
SN~2001ex & UGC~3595                & 35.24 (0.15) & HF      & 20      & 52204.80 (0.24) & 17.324 (0.041) & 1.695 (0.099) & \phs1.006 (0.083) & 0.415 (0.148) & 23       \\
SN~2003bf & 2MASX~J08082660+1219571 & 35.81 (0.15) & HF      & 20      & \nodata         & \nodata        & \nodata       & \nodata           & \nodata       & \nodata  \\
SN~2003bt & MCG~-01-28-006          & 35.34 (0.15) & HF      & 20      & \nodata         & \nodata        & \nodata       & \nodata           & \nodata       & \nodata  \\

\tablerefs{(1) \citealt{Jensen03}; (2) \citealt{Hamuy91};
(3) \citealt{Freedman01}; (4) \citealt{Buta83};
(5) \citealt{Tsvetkov82}; (6) \citealt{Ciatti88};
(7) \citealt{Tonry01}; (8) \citealt{Buta85}; (9) \citealt{Younger85};
(10) \citealt{Phillips87}; (11) \citealt{Wells94};
(12) \citealt{Lira98}; (13) \citealt{Gibson00}; (14) \citealt{Lira98};
(15) \citealt{Altavilla04}; (16) \citealt{Filippenko92:91bg};
(17) \citealt{Leibundgut93}; (18) \citealt{Turatto96};
(19) N. Suntzeff 2005, private communication;
(20) NASA/IPAC Extragalactic Database;
(21) K. Krisciunas 2005, private communication;
(22) \citealt{Krisciunas04}; (23) \citealt{Ganeshalingam08};
(24) \citealt{Krisciunas03}.}

\tablecomments{Uncertainties are given in parentheses.}

\tablenotetext{a}{Method of determining distance modulus --- C =
Cepheid variable stars; SBF = surface brightness fluctuations; 
HF = Hubble flow.}

\enddata

\end{deluxetable}
\end{center}

\clearpage

\LongTables
\tabletypesize{\footnotesize}
\begin{deluxetable}{llcrccc}
%\rotate
\tablewidth{0pt}
\tablecaption{Spectroscopic Information\label{t:spec}}
\tablehead{
\colhead{SN} &
\colhead{UT Date} &
\colhead{Telescope /} &
\colhead{Phase} &
\colhead{\ion{Fe}{2} Velocity} &
\colhead{\ion{Fe}{2} EW} &
\colhead{UV Ratio} \\
\colhead{Name} &
\colhead{yyyy-mm-dd} &
\colhead{Instrument} &
\colhead{(d)} &
\colhead{($10^{3}$ km~s$^{-1}$)} &
\colhead{(\AA)} &
\colhead{$\mathcal{R}_{UV}$}}

\startdata

SN~1980N  & 1980-12-11.724 & {\it IUE}      &  $-0.2$ & 18.2    & 58.2    & 0.28 (0.02) \\
SN~1980N  & 1980-12-13.495 & {\it IUE}      &    1.6  & 21.6    & 69.6    & 0.29 (0.03) \\
SN~1980N  & 1980-12-16.147 & {\it IUE}      &    4.2  & 16.8    & 80.3    & 0.34 (0.04) \\
SN~1980N  & 1980-12-17.712 & {\it IUE}      &    5.8  & 23.5    & 10.2    & 0.41 (0.04) \\
SN~1980N  & 1980-12-23.371 & {\it IUE}      &   11.4  & 23.5    & 17.0    & 0.58 (0.05) \\
SN~1980N  & 1980-12-24.831 & {\it IUE}      &   12.9  & 24.5    & 11.8    & 0.59 (0.04) \\
SN~1980N  & 1981-01-16.810 & {\it IUE}      &   35.7  & 22.0    & 55.9    & 0.90 (0.09) \\
SN~1981B  & 1981-03-09.295 & {\it IUE}      &    1.8  & 20.6    & 47.3    & 0.32 (0.01) \\
SN~1981B  & 1981-03-10.460 & {\it IUE}      &    3.0  & 21.4    & 49.0    & 0.27 (0.02) \\
SN~1981B  & 1981-03-11.597 & {\it IUE}      &    4.1  & \nodata & \nodata & 0.53 (0.06) \\
SN~1982B  & 1982-02-18.436 & {\it IUE}      &    6.0  & 20.2    & 27.2    & \nodata     \\
SN~1983G  & 1983-04-08.481 & {\it IUE}      &    3.2  & 20.1    & 44.0    & 0.53 (0.03) \\
SN~1983G  & 1983-04-09.368 & {\it IUE}      &    4.1  & \nodata & \nodata & 0.45 (0.15) \\
SN~1983G  & 1983-04-11.544 & {\it IUE}      &    6.3  & 19.5    & 20.5    & 0.44 (0.03) \\
SN~1983G  & 1983-04-14.281 & {\it IUE}      &    9.0  & 20.8    & 56.3    & 0.56 (0.02) \\
SN~1983G  & 1983-04-19.569 & {\it IUE}      &   14.3  & 19.7    & 56.7    & 0.27 (0.05) \\
SN~1983G  & 1983-04-25.289 & {\it IUE}      &   20.0  & 21.7    & 67.9    & 0.54 (0.03) \\
SN~1986G  & 1986-05-06.685 & {\it IUE}      &  $-4.2$ & \nodata & \nodata & 0.51 (0.18) \\
SN~1986G  & 1986-05-08.146 & {\it IUE}      &  $-2.8$ & 28.6    & 42.9    & 0.45 (0.08) \\
SN~1986G  & 1986-05-12.493 & {\it IUE}      &    1.6  & 21.4    & 71.1    & 0.39 (0.02) \\
SN~1986G  & 1986-05-15.138 & {\it IUE}      &    4.2  & \nodata & \nodata & 0.46 (0.03) \\
SN~1986G  & 1986-05-19.321 & {\it IUE}      &    8.4  & 25.1    & 47.7    & 0.42 (0.09) \\
SN~1986G  & 1986-05-29.301 & {\it IUE}      &   18.3  & 20.9    & 79.2    & 0.88 (0.08) \\
SN~1989B  & 1989-02-01.480 & {\it IUE}      &  $-5.3$ & 20.2    & 19.7    & 0.43 (0.01) \\
SN~1989M  & 1989-07-06.988 & {\it IUE}      & \nodata & 20.0    & 62.4    & \nodata     \\
SN~1989M  & 1989-07-08.569 & {\it IUE}      & \nodata & 19.5    & 49.5    & \nodata     \\
SN~1989M  & 1989-07-12.735 & {\it IUE}      & \nodata & 18.8    & 42.0    & \nodata     \\
SN~1989M  & 1989-07-14.058 & {\it IUE}      & \nodata & 17.7    & 34.6    & \nodata     \\
SN~1989M  & 1989-07-17.730 & {\it IUE}      & \nodata & 15.1    & 39.8    & \nodata     \\
SN~1989M  & 1989-07-20.034 & {\it IUE}      & \nodata & 16.9    & 37.8    & \nodata     \\
SN~1990M  & 1990-06-15.836 & {\it IUE}      & \nodata & 23.1    & 24.5    & \nodata     \\
SN~1990M  & 1990-06-18.493 & {\it IUE}      & \nodata & 23.0    & 16.9    & \nodata     \\
SN~1990N  & 1990-06-26.817 & {\it IUE}      & $-12.5$ & \nodata & \nodata & 0.41 (0.02) \\
SN~1990N  & 1990-06-28.391 & {\it IUE}      & $-11.0$ & 28.8    & 63.6    & 0.60 (0.06) \\
SN~1990N  & 1990-06-30.726 & {\it IUE}      &  $-8.6$ & 27.5    & 37.9    & 0.41 (0.01) \\
SN~1990N  & 1990-07-02.727 & {\it IUE}      &  $-6.6$ & 30.8    & 30.6    & 0.35 (0.03) \\
SN~1990N  & 1990-07-04.333 & {\it IUE}      &  $-5.0$ & 32.6    & 42.2    & 0.35 (0.01) \\
SN~1990N  & 1990-07-09.730 & {\it IUE}      &    0.3  & 24.2    & 17.4    & 0.29 (0.01) \\
SN~1990N  & 1990-07-14.301 & {\it IUE}      &    4.9  & 21.6    & 23.9    & 0.37 (0.01) \\
SN~1991T  & 1991-04-27.424 & {\it IUE}      &  $-0.1$ & 22.3    & 16.9    & 0.26 (0.01) \\
SN~1991T  & 1991-04-29.672 & {\it IUE}      &    2.1  & 21.7    & 20.7    & 0.23 (0.01) \\
SN~1991bg & 1991-12-14.824 & {\it IUE}      &    2.4  & \nodata  & \nodata & \nodata     \\
SN~1992A  & 1992-01-14.083 & {\it IUE}      &  $-5.0$ & 20.4    & 45.1    & 0.33 (0.01) \\
SN~1992A  & 1992-01-16.220 & {\it IUE}      &  $-2.9$ & 20.7    & 49.7    & 0.41 (0.03) \\
SN~1992A  & 1992-01-18.952 & {\it IUE}      &  $-0.2$ & 19.2    & 51.5    & 0.36 (0.02) \\
SN~1992A  & 1992-01-21.574 & {\it IUE}      &    2.4  & 18.6    & 61.4    & 0.37 (0.01) \\
SN~1992A  & 1992-01-24.214 & {\it HST}/FOS  &    5.1  & 19.1    & 55.0    & 0.45 (0.03) \\
SN~1992A  & 1992-01-27.666 & {\it IUE}      &    8.5  & 21.4    & 49.0    & 0.53 (0.02) \\
SN~1992A  & 1992-01-30.906 & {\it IUE}      &   11.7  & 21.5    & 61.3    & 0.60 (0.01) \\
SN~1992A  & 1992-02-04.218 & {\it IUE}      &   16.0  & \nodata & \nodata & 0.78 (0.03) \\
SN~1992A  & 1992-02-10.659 & {\it IUE}      &   22.4  & \nodata & \nodata & 1.08 (0.12) \\
SN~1992A  & 1992-03-04.310 & {\it HST}/FOS  &   44.9  & \nodata & \nodata & 0.87 (0.02) \\
SN~1992A  & 1992-11-05.592 & {\it HST}/FOS  &  289.7  & \nodata & \nodata & 0.55 (0.06) \\
SN~2001ay & 2001-05-02.813 & {\it HST}/STIS &    8.3  & 19.3    & 21.5    & \nodata     \\
SN~2001ay & 2001-05-09.704 & {\it HST}/STIS &   15.0  & 19.2    & 30.1    & \nodata     \\
SN~2001ba & 2001-05-08.791 & {\it HST}/STIS &    4.0  & 19.3    & 48.5    & \nodata     \\
SN~2001ba & 2001-05-16.229 & {\it HST}/STIS &   11.2  & 18.0    & 47.4    & \nodata     \\
SN~2001eh & 2001-09-25.757 & {\it HST}/STIS &    8.1  & \nodata & \nodata & 0.33 (0.01) \\
SN~2001eh & 2001-10-04.523 & {\it HST}/STIS &   16.6  & 26.4    & 48.3    & 0.49 (0.02) \\
SN~2001el & 2001-10-29.636 & {\it HST}/STIS &   29.9  & 18.3    & 28.8    & \nodata     \\
SN~2001el & 2001-11-06.569 & {\it HST}/STIS &   37.8  & 17.2    & 39.9    & \nodata     \\
SN~2001el & 2001-11-13.476 & {\it HST}/STIS &   44.7  & 15.9    & 42.4    & \nodata     \\
SN~2001el & 2001-11-20.356 & {\it HST}/STIS &   51.5  & 18.6    & 46.2    & \nodata     \\
SN~2001el & 2001-11-26.704 & {\it HST}/STIS &   57.8  & 18.6    & 40.9    & \nodata     \\
SN~2001el & 2001-12-04.350 & {\it HST}/STIS &   65.5  & 15.5    & 55.2    & \nodata     \\
SN~2001ep & 2001-10-28.024 & {\it HST}/STIS &   10.3  & \nodata & \nodata & \nodata     \\
SN~2001ep & 2001-11-02.871 & {\it HST}/STIS &   16.1  & \nodata & \nodata & \nodata     \\
SN~2001ex & 2001-10-29.475 & {\it HST}/STIS &    7.0  & 23.1    & \nodata & \nodata     \\
SN~2001ex & 2001-11-06.424 & {\it HST}/STIS &   14.7  & \nodata & \nodata & \nodata     \\
SN~2001ex & 2001-11-12.786 & {\it HST}/STIS &   20.9  & \nodata & \nodata & \nodata     \\
SN~2001ex & 2001-11-19.272 & {\it HST}/STIS &   27.3  & \nodata & \nodata & \nodata     \\
SN~2001ex & 2001-11-26.885 & {\it HST}/STIS &   34.7  & \nodata & \nodata & \nodata     \\
SN~2001ex & 2001-12-03.964 & {\it HST}/STIS &   41.6  & \nodata & \nodata & \nodata     \\
SN~2003bf & 2003-03-10.836 & {\it HST}/STIS & \nodata & 17.0    & 42.6    & \nodata     \\
SN~2003bf & 2003-03-17.577 & {\it HST}/STIS & \nodata & \nodata & \nodata & \nodata     \\
SN~2003bf & 2003-03-24.245 & {\it HST}/STIS & \nodata & \nodata & \nodata & \nodata     \\
SN~2003bf & 2003-03-17.713 & {\it HST}/STIS & \nodata & \nodata & \nodata & \nodata     \\

\enddata

\tablecomments{Uncertainties are given in parentheses.}

\end{deluxetable}

%\eject

\end{document}